\documentclass[prl,twocolumn, amsmath,amssymb]{revtex4}
\usepackage[utf8]{inputenc}
\usepackage[english]{babel}
\usepackage[dvips]{graphics}
\usepackage{xcolor}
\newcommand{\qed}{\hfill \mbox{\raggedright \rule{.07in}{.1in}}}

\newcommand{\ket}[1]{\left | #1 \right\rangle}

\newcommand{\braket}[2]{\left\langle #1|#2\right\rangle}

\def\identity{\leavevmode\hbox{\small1\kern-3.8pt\normalsize1}}

\begin{document}

\title{The information-theoretic foundation of thermodynamic work extraction}

\author{Chiara Marletto}

\affiliation{Clarendon Laboratory, University of Oxford, UK\\Centre for Quantum Technologies, National University of Singapore, Singapore}

\date{\today}

\begin{abstract}
{In this paper I apply newly-proposed information-theoretic principles to thermodynamic work extraction. I show that if it is possible to extract work deterministically from a physical system prepared in any one of a set of states, then those states must be distinguishable from one another.  This result is formulated independently of scale and of particular dynamical laws; it also provides a novel connection between thermodynamics and information theory, established via the law of conservation of energy (rather than the second law of thermodynamics). Albeit compatible with these conclusions, existing thermodynamics approaches cannot provide a result of such generality, because they are scale-dependent (relying on ensembles or coarse-graining) or tied to particular dynamical laws. This paper thus provides a broader foundation for thermodynamics, with implications for the theory of von Neumann's universal constructor.}

\noindent 

\end{abstract}

\maketitle             

Microscopic dynamical laws are time-reversal symmetric. Hence the second law of thermodynamics, intended as mandating the irreversibility of certain dynamical trajectories, is ruled out at the microscopic scale. This tension is usually tackled with statistical mechanics: Boltzmann's and Gibbs' ensemble theories, \cite{UFF}, and their quantum-mechanical generalisations in the hotly investigated area of quantum thermodynamics \cite{THD1, THD2, THD3}. These powerful methodologies derive the second law from classical or quantum dynamics with additional assumptions. 

Despite their tremendous success in a broad set of regimes, these schemes have problems at their foundations. First, some such schemes traditionally rely on approximations such as ensembles and coarse-graining, which make the ensuing second laws {\sl scale-dependent}, \cite{UFF}, and only applicable at a certain macroscopic scale, which is never exactly defined. Examples of scale-dependent laws are those about ferromagnetic phase transitions,  which become exact only in the thermodynamic limit (and are not even intended to be exact for realistic systems). I shall designate as {\sl `scale-independent'} any law whose applicability to a system does not depend on the system's scale.  Most fundamental laws are scale-independent, e.g. conservation laws or Einstein's equations.

Furthermore, some formulations of the second law are tied to a particular class of dynamical laws: for instance, quantum thermodynamics is formulated within quantum theory. Hence, they are less general than traditional thermodynamics, which consists of a set of meta-laws largely independent of the details of the dynamical laws they constrain. I shall call laws which can be expressed without reference to the details of any particular dynamics, {\sl `dynamics-independent'} \footnote{The term `dynamics' here refers to a law of motion, including both formal kinematic elements (e.g. the algebra of observables in quantum theory) and dynamical ones (e.g. the equations of motion)}.
Here I propose a new information-theoretic foundation for thermodynamic laws which is independent of scale (refers to no particular length or time or complexity) and of dynamics (i.e. refers to no particular equations of motion).

The key result is that under a set of general principles (which are satisfied by quantum theory and by all physical theories that are seriously contemplated at present -- but also by a vastly larger class):\\

{{\sl If it is possible to extract thermodynamic work deterministically from a physical system prepared in any one of a set of attributes, then the attributes in that set are all distinguishable from one another,}}
\\

\noindent {where `attribute' hereinafter indicates a set of states. Crucially, I shall define `extracting thermodynamic work' and `distinguishable' in a {\sl scale-independent} and {\sl dynamics-independent} way. Definitions of these concepts already exist, expressed within particular dynamics. For instance: in quantum information, two qubit states are distinguishable if and only if they are orthogonal; in quantum thermodynamics the work deterministically extractable, asymptotically, in a process taking a quantum state $\rho_1$ to $\rho_2$ is given by: $F(\rho_1)-F(\rho_2)$, where $F(\rho)=U(\rho)-\kappa_BTS(\rho)$, and $S(\rho)=-{\rm Tr}\{\rho\ln{\rho}\}$ while $U(\rho)= {\rm Tr}\{\rho H\}$, $H$ being the Hamiltonian of the isolated system. These propositions are formulated using quantum theory's formalism, hence they are dynamics-dependent. My results will be consistent with these dynamics-dependent notions of distinguishability and work extraction,  but formulated in a strictly scale- and dynamics-independent way, thus being more general. To this end I shall assume the principles of the recently proposed {\sl constructor theory of information} \cite{DEU, DEUMA} -- consisting of new scale-independent, dynamics-independent physical principles.}

Specifically, I shall propose a scale-independent, dynamics-independent definition of thermodynamic {\sl work extraction}. It includes as special cases the classical and quantum-thermodynamics definitions, but it is more general. I shall also establish a further unexpected connection between thermodynamics and information theory, by showing that the possibility of extracting work {\sl deterministically} from a system prepared in any one of a set of states implies that those states must all be distinguishable (in the information-theoretic sense, which is far more general than the quantum one) from one another. Surprisingly, this link between information theory and thermodynamics goes via the law of {\sl conservation of energy}, instead of the second law. This result poses a fundamental limitation on any quantum thermodynamics protocol for extracting work from systems with quantum coherence, e.g. \cite{THD1, JEN1}.

\noindent {\bf Constructor theory of information.} I now summarise informally the basics of constructor theory (CT) (see appendix A and \cite{DEUMA, MAP, MAT} for the formal details). The fundamental concept in CT is that of a task. A task is the specification of a transformation expressed as an ordered pair of input/output attributes. Attributes are sets of states of a physical system on which tasks can be performed, which are called `substrates'.  If ${\bf a}$ and ${\bf b}$ are attributes, the attribute ${\bf (a,b)}$ of the composite system ${\bf S_1}\oplus {\bf S_2}$ is defined as the set of all states of the composite system where ${\bf S_1}$ has attribute ${\bf a}$ and ${\bf S_2}$ has attribute ${\bf b}$. In quantum theory, for instance, a qubit is a substrate; one of its attributes is a set of states such that a given projector is sharp with value 1 in each state of that set. Denoting by ${\bf 0}$ the attribute for the qubit's state to be in a given subspace and by ${\bf 1}$  the attribute for the qubit's state to be in its orthogonal complement, an example of a task is $\{{\bf 0} \rightarrow{\bf 1},{\bf 1} \rightarrow{\bf 0}\}$, negating the qubit in a particular basis. A {\sl variable} is a set of disjoint attributes. Given a task $T$, define its {\sl transpose} as the task obtained from $T$ by swapping each input attribute with the corresponding output attribute. 

A {\sl constructor} for a task $T$ is a system which whenever presented with the substrate of $T$ in any state belonging to one of the input attributes, it delivers it in one of the states of the allowed output attributes, and {\sl retains the ability to do that again}, \footnote{The notion of a catalyst in resource theory \cite{LAN} could be considered as a model for special cases of constructors - catalysts must stay in exactly the same state (as opposed to the same attribute)  and their definition is dynamics-dependent.}. In quantum theory (see appendix A), a constructor is modelled by a subspace $C$ with the following property: the substrate undergoes the transformation specified by $T$, whenever it is coupled to the environment in a state belonging to $C$, and $C$ is invariant under the action of the overall unitary evolution of the joint system of substrates and environment. 

 \noindent A task is {\sl impossible} if the laws of physics impose a limit on how accurately it can be performed by a constructor. Otherwise, the task is {\sl possible}. In quantum information, gates are example of constructors \cite{DEU}: logically reversible computational tasks are all possible under the unitary quantum model of quantum computation.  {CT consists of general newly-conjectured principles expressed solely in terms of possible/impossible tasks, intended to supplement laws of motion (such as quantum theory's or general relativity's), which are called {\sl subsidiary theories}. The full explanation of a given physical situation is given by the principles of CT and by the compatible subsidiary theories. The principles are formulated in a scale- and dynamics-independent way, so they underlie a number of subsidiary theories; they don't refer to constructors, rather to the possibility or impossibility of certain tasks. Here I shall confine attention to subsidiary theories with a space of allowed states endowed with a topology assigning a meaning to states being arbitrarily close to each other. {For present purposes it is not necessary to model a possible task within a given subsidiary theory, because I shall take it as primitive, just like in the theory of computation. I discuss a simple quantum model in appendix A, following \cite{THE}.}

\noindent {\bf Distinguishability.} 

The base of my construction is a scale-independent, dynamics-independent definition of distinguishability, \cite{DEUMA}. It generalises the quantum-information notion of states that can be distinguished arbitrarily well from each other with a single- shot, projective measurement (without referring to orthogonality). 

First one defines a class of substrates, {\sl information media}, by requiring that some tasks are possible on them - tasks that are conjectured to be sufficient for them to be capable of carrying information. In short, information media must have a variable $X$ with the property that it is possible to perform all the permutation tasks on X, and that it is possible to perform the task of copying all attributes in $X$ from one substrate to its replica (see appendix A for the formal definitions). 

Any variable $X$ that can be copied and permuted in all possible ways is called an {\sl information variable}. An example of information medium is a qubit with an information variable being any set of two orthogonal states. 

\noindent Any two different information media (e.g. a neutron and a photon) must satisfy an {\sl interoperability principle}, \cite{DEUMA}, which expresses elegantly the intuitive property that classical information must be copiable from one information medium to any other (having the same capacity), irrespective of their physical details. Specifically, if $S_1$ and $S_2$ are information media, respectively with information variable $X_1$ and $X_2$, their composite system $S_1 \oplus S_2$ is an information medium with information variable $X_1\times X_2$, where $\times$ denotes the Cartesian product of sets. 

Now I define distinguishability using information media, as follows. A variable $Y$ is {\sl distinguishable} if (informally) it is physically possible to map it onto  an information variable in a logically reversible fashion, i.e. if the task

\begin{equation}
\label{eq3}
\bigcup_{{y}\in Y}\left\{{\bf y}\rightarrow {\bf q_y} \right\}
\end{equation}

\noindent is possible, where the variable $\{{\bf q_y}\}$, of the same cardinality as $Y$, is an information variable. {Hence, a set of orthogonal quantum states for which the above task is possible is a distinguishable variable - but we have expressed this fact without referring to quantum theory's specific formalism, in a scale- and dynamics-independent way. }

A principle of CT that I shall deploy is the {\sl principle of asymptotic distinguishability}. Informally, it requires that $N$ copies of an attribute ${\bf x}$, and $N$ copies of another attribute that is disjoint from ${\bf x}$ are asymptotically distinguishable. In other words, the task of distinguishing them is possible as the number $N$ of copies goes to infinity (where e.g. having one of a specified set of density matrices counts as an attribute). In quantum theory, this corresponds to the fact that any two different quantum states are tomographically distinguishable.

\noindent {\bf Work media}. In traditional thermodynamics there is a general consensus, following Planck, on identifying a work repository with a system behaving `in the same way' as a weight in a uniform gravitational field, which can be smoothly raised or lowered to different heights, \cite{UFF}.  In quantum thermodynamics, it is common practice to define a work repository as a system in any eigenstate of its free Hamiltonian, such as a set of bound states in an atom, utilisable as a battery; there are also other proposed notions of work repositories (see \cite{THD2} for a review). Here my intention is to be more general than those notions, but compatible with all of them. I shall do so by generalising the class of work repositories to that of {\sl work media}, \cite{MAT}.  I shall define work media as a particular class of substrates satisfying an operational criterion (just like information media): certain tasks must be possible on a substrate for it to qualify as a work medium. This will provide a conjectured scale- and dynamics-independent generalisation of the notion of work repository, building on the classical definition of Planck's and Clausius'. 

First, one needs to express the conservation of energy in CT. Following \cite{DEU}, it is possible to show that the presence of a conservation law implies that tasks on a given substrates are partitioned into {\sl equivalence classes}. Here I shall call these classes `energy-equivalence-classes', as I will focus on energy conservation only. Tasks belonging to the same equivalence class violate the energy conservation by the same amount - see appendix B for details. 

{ A work medium is a substrate ${\bf Q}$ having a variable $W=\{{\bf w_+}, {\bf w_0}\}$ with the property that:}

\begin{itemize}

\item{} The task $T_{+,0}=\{{\bf w_+}\rightarrow{\bf w_0}\}$ belongs to an energy-equivalence class such that $T_{+,0}$ is impossible and so is its transpose. 
\item{} There exists an attribute ${\bf w_-}$ of ${\bf Q}$, disjoint from ${\bf w_0}$ and ${\bf w_+}$, such that the task:
\begin{equation}
\{({\bf w_+},{\bf w_0}) \rightarrow ({\bf w_0},{\bf w_+}), ({\bf w_0},{\bf w_0}) \rightarrow ({\bf w_+},{\bf w_-})\} \;\;\label{SWAP}
\end{equation}

is possible.

\end{itemize}

\noindent Such a variable $W$ is a {\sl work variable} {\footnote{Other tasks that are not specified above may or may not be possible, depending on the subsidiary theory.}}.

{ An example of a system possessing a work variable is an atom $Q$ with three different equally-spaced energy levels, in decreasing order of energy as follows: ${\bf w_+}, {\bf w_0}, {\bf w_-}$.} {In the presence of finite resources, it is impossible to perform the task $T_{+,0}$, because of the conservation of energy: the task requires the energy of the atom to change. For, due to energy conservation, any finite-dimensional environment coupled to the atom would have to modify its energy by an amount that is equal and opposite to the amount by which $T_{+,0}$ changes the energy of the atom, hence it cannot act as a constructor for the task. Thus condition (i) is satisfied.  Finally, the task in \eqref{SWAP} is possible, by a suitably engineered unitary that is energy-preserving. So, a quantum system with at least 3 equally spaced energy levels satisfies the definition of work media, hence this definition is compatible with existing classical and quantum notions of work repository.}

{The key fact about condition (ii) (that the task \eqref{SWAP} is possible) is that it is {\sl not} satisfied by purely thermal attributes such as having a particular temperature, in line with traditional thermodynamics: as is well-known, a single thermal state cannot be used to do work.  For example, let's assume ${\bf w_{\alpha}}={\bf T_{\alpha}}$, where the attributes ${\bf T_+}, {\bf T_-}, {\bf T_0}$ of, say, a volume of water each correspond to a thermal state with given temperature $T_{\alpha}$. In order to satisfy the first requirement (equation \eqref{SWAP}),  an equilibrium state $({\bf T_0},{\bf T_0})$ should be transformed into the temperature attribute $({\bf T_+},{\bf T_-})$, with no other side effects. This is impossible according to the second law in classical and quantum thermodynamics. Thus, systems endowed with thermal degrees of freedom within the standard definitions of thermodynamics do not qualify as work media.}

The above definition identifies precisely the attributes that can be used to acquire energy from another system, or deliver energy to it, {\sl reversibly}, with no other side-effects. It is consistent with the traditional notion of `work repository' or `mechanical means', but it is applicable to general systems that need not be mechanical, e.g. an atom in an excited state. It advances existing definitions, such as those declaring eigenstates of energy to be work repositories by fiat. So it is a good candidate to use in order to build a scale- and dynamics-independent notion of deterministic work extraction. {Note also that this is a set of sufficient, operational conditions for a physical system to behave like a work repository. There could be tighter definitions, but for present purposes we only need to consider sufficient conditions.}

\noindent {\bf A deterministic work extractor.} The task of deterministically extracting work from a substrate {\bf S} in regard to a variable X of {\bf S} is defined as:

\begin{equation}
\bigcup_{x\in X}\{({\bf x},{\bf w_0}) \rightarrow ({\bf f_x},{\bf w_x})\} \label{REF}
\end{equation}

\noindent where $\{{\bf f_x}\}$ is some variable of ${\bf S}$ and the pairs $\{{\bf w_x},{\bf w_{x'}}\}$, for all $x, x'\in X$, are each a work variable of ${\bf M}$. For example, ${\bf M}$ here could be an atom with several levels of energy that gets excited or de-excited by interaction with another system ${\bf S}$.
A constructor for the above task is deterministic because it delivers with certainty one and only one output attribute for any particular input attribute, retaining the ability to do that again, and without any other side-effects. { Such reliable behaviour is expected of an ideal classical heat engine and of an ideal quantum deterministic work extractor, \cite{THD1}, so this requirement is well-grounded in existing theories of thermodynamics. By continuity, one could also consider probabilistic work extractors, in which case what follows would still hold, with a certain probability set by the reliability of the probabilistic work extractor. Investigating the probabilistic case is outside of the scope of this paper.}

\noindent {\bf The information-theoretic foundation of deterministic work extraction}. I can now state the key result of the paper more formally:
\\
{\center{{\bf Theorem 1.} {\sl A work variable is a distinguishable variable.} }}
\\

{\noindent This follows straightforwardly from the fact that the task \eqref{SWAP} is possible on a work medium.} Consider a work variable $W$ and the following task, generalising \eqref{SWAP} to having $n$ substrates as target:

\begin{eqnarray}
\{({\bf w_+},({\bf w_0})^{(2n)} )\rightarrow ({\bf w_+}, {({\bf w_+}, {\bf w_-})}^{ (n)});\nonumber \\ 
({\bf w_0},({\bf w_0})^{(2n)} )\rightarrow ({\bf w_0}, {({\bf w_-}, {\bf w_+})}^{ (n)})\} \label{X}.
\end{eqnarray}

\noindent When $n$ tends to infinity, ${({\bf w_+}, {\bf w_-})}^{ (n)})$ is asymptotically distinguishable from $ {({\bf w_-}, {\bf w_+})}^{ (n)}$, by the asymptotic-distinguishability principle. Thus, the attributes ${\bf w_+}$ and ${\bf w_0}$ of a work medium are distinguishable from one another, by definition of distinguishability. {The proof is expressible in quantum theory, by modelling the attributes as non-intersecting linear subspaces and using standard results from state tomography (see \cite{THE}, a summary of which is in appendix C).} 

Hence, by this proof, any variable $X$ for which the task \eqref{REF} of deterministically extracting work is possible must also be distinguishable. 
This concludes the proof that a deterministic work extractor is also a perfect distinguisher  -- hence all the attributes in a work variable (from which work can be extracted realiably) must be distinguishable from one another. \noindent Specialising to quantum theory, this implies that work can only be extracted from a system prepared in one of a set of orthogonal subspaces.

\noindent {\bf A new scale- and dynamics-independent foundation for the second law.}  {This theorem tackles the issue of formulating the second law of thermodynamics in a scale- and dynamics-independent way. I can illustrate how by recalling the concept of adiabatic accessibility (epitomised by the famous Joule's experiments, \cite{UFF}) -- the core of the axiomatic approach to thermodynamics, \cite{CAR, LIE, BUC, UFF}. An attribute ${\bf b}$ is adiabatically accessible from the attribute ${\bf a}$ if it is possible to construct a thermodynamic cycle that transforms ${\bf a}$ into ${\bf b}$ with the only side-effect being the raising or lowering of a weight in a gravitational field. So for instance the second law in traditional thermodynamics says that the state of a volume of water at a given temperature is adiabatically accessible from one at a lower temperature (because mechanical stirring can heat up an otherwise isolated volume of water); but it is not adiabatically accessible from a state at a higher temperature (because mechanical stirring cannot by itself cool an otherwise isolated volume of water).}  

Using work media, one can propose a variant of the definition of adiabatic accessibility, appealing to the notion of {\sl adiabatic possibility}, with the crucial advantage of being scale- and dynamics-independent. A task $\{ {\bf x}\rightarrow {\bf y}\}$ is adiabatically possible if the task: 

 $$
 \{({\bf x},{\bf w_1})\rightarrow ({\bf y},{\bf w_2}) \}
 $$

is possible for some two work attributes ${\bf w_1}, {\bf w_2}$ belonging to a work variable. The latter generalises the ad-hoc weight-in-a-gravitational-field criterion invoked in the traditional definition, making the notion of adiabatic accessibility dynamics-independent (as the definition of work media is also dynamics-independent). Also, this definition does not depend on coarse-graining, so it is scale-independent. Therefore it allows one to formulate a dynamics- and scale-independent {\sl second law}, \cite{MAT}, expressed as the requirement that:

{\center \sl{There are tasks that are adiabatically possible, whose transpose is not adiabatically possible.}} \\

\noindent Note how the statement is fully compatible with traditional thermodynamics laws, but it extends as it is scale- and dynamics- independent. This provides the basis for a scale-independent distinction between work and heat.

\noindent   {\bf Discussion.} My theorem establishes a novel foundation for thermodynamics, based on constructor-information theory, which is scale- and dynamics-independent. In quantum theory, this result implies that if one can extract work deterministically from any of a set of states, these states must be orthogonal to each other. The theorem I proved is similar in logic to the no-cloning theorem in quantum information: it is a no-go theorem, stating that one cannot extract work reliably from a system prepared in any one of a set of states unless they are perfectly distinguishable. However, it is more far-reaching than the no-cloning theorem, because it is dynamics-independent, so it is more general than, but compatible with, quantum theory. For instance, it could apply to the potential successors of quantum theories - e.g. theories of coupled gravity and quantum matter. It therefore provides a promising basis for constraining future subsidiary theories, including those describing exotic objects such as black holes or closed time-like curves. 
It also connects information theory and thermodynamics in an unexpected way, not regarding the second law, but the conservation of energy. 

An interesting parallel between a programmable quantum computer (whose admissible programs must belong to the computational basis \cite{MYE, NIE}) and a deterministic work extractor emerges here. I proved that variables that can serve as input to a deterministic work extractor must be a set of distinguishable attributes. This constitutes the only possible `work basis', which, like the computational basis, has to consist of distinguishable, orthogonal subspaces. These could be either a set of sharp energy states; or a set of states that are not diagonal in the energy basis, each provided with orthogonal labels. Hence, it is impossible to extract work deterministically in a single-shot fashion from a set of unknown (pure or mixed) quantum states with a given average work content, e.g. states produced by a naturally occurring phenomenon. This poses a fundamental limit on the work that can be extracted deterministically from quantum systems with coherence in the energy basis. One can envisage a process that extracts work optimally and deterministically from a particular, known, quantum state with some non-zero coherence in the energy basis, \cite{JEN1}, as compared to the corresponding thermal state with the same mean energy. However, this process is a special-purpose machine, which requires to know a priori which state has been prepared. Therefore, it is not a universal work-extractor in the conventional thermodynamic sense, not more than a Szilard engine without its memory is. 

This work provides the foundation for formulating thermodynamics in an information-theoretic, dynamics-independent and scale-independent way: hence, it can inform new experimental schemes to test this proposed scale- and dynamics-independent reformulation of the second law, see e.g. \cite{GEN}. It is also a first step towards a theory of programmable constructors in quantum theory, which will generalise the theory of quantum computation to general tasks, in a way already envisaged in von Neumann' theory of the universal constructor. In order to devise this theory, one will have to merge quantum thermodynamics with general principles of CT. \\{\bf Acknowledgements} The author thanks David Deutsch and Vlatko Vedral for discussions and comments on earlier versions of this manuscript; Benjamin Yadin and Paul Raymond-Robichaud for helpful suggestions.\\

\subsection{Appendix A: Constructor Theory} %

Constructor theory is a meta-theory with its own physical principles that are intended to supplement and constrain dynamical theories, such as quantum theory and general relativity, which therefore we call {\sl subsidiary theories}, \cite{DEU}. Every subsidiary theory that is constructor-theory compliant must provide a set of allowed states of the allowed substrates, endowed with a topology. 

An {\sl attribute} ${\bf x}$ is a set of states all having a property $x$.  For instance, in quantum theory, the set of all quantum states of a qubit where a given projector $\Pi$ is sharp with value $1$ is an attribute.  A {\sl variable} is a set of disjoint attributes. 

If ${\bf a}$ is an attribute of substrate ${\bf S_1}$ and ${\bf b}$ is an attribute of substrate ${\bf S_2}$, the attribute ${\bf (a,b)}$ of the composite substrate ${\bf S_1}\oplus {\bf S_2}$ is defined as the set of all states where ${\bf S_1}$ has attribute ${\bf a}$ and ${\bf S_2}$ has attribute ${\bf b}$. Einstein's principle of locality requires that if a transformation operates only on substrate ${\bf S_1}$, then only the attribute ${\bf a}$ changes, not ${\bf b}$, \cite{DEUMA}. 

In quantum theory, assuming a two-qubit Hilbert space ${\cal H}_{ab}$, $${\bf (a,b)}\doteq\{ \rho_{ab}\in {\cal H}_{ab}\;:\;{\rm Tr}\{\rho_{ab}\Pi_a\otimes \Pi_b \}=1\}$$ where $\Pi_a$ and $\Pi_b$ are given projectors defined on each qubit's Hilbert space. Note that this attribute may include quantum states where the qubits are entangled. 

A {\sl task} is the abstract specification of a physical transformation, represented as a finite set of ordered pairs of input/output attributes: $T=\{{\bf a_1}\rightarrow{\bf b_1}, {\bf a_2} \rightarrow {\bf b_2},\;\cdots\;, {\bf a_n}\rightarrow{\bf b_n}\}$. 

\noindent A {\sl constructor} for a task $T$ is a system which whenever presented with the substrate of the task $T$ in one of the input attributes, it delivers it in one of the states of the allowed output attributes, and {\sl retains the ability to do that again}. A task is {\sl impossible} if the laws of physics impose a limit on how accurately it can be performed by a constructor. Otherwise, the task is {\sl possible}. Constructor-theoretic statements never refer to specific constructors, only to the fact that tasks are possible or impossible. This is what allows them to be scale- and dynamics-independent.

Tasks close an algebra, \cite{THE}. Two tasks $T_1$ and $T_2$ can be composed in series (whenever the output set of attributes of $T_1$ includes the input set of attributes of $T_2$), or in parallel, with the usual informal meaning of parallel and serial composition, {\cite{DEUMA}}. I denote the serial composition of two tasks as $T_1T_2$; the parallel composition as $T_1\otimes T_2$. The transpose of a task $T$, denoted by $T^{\sim}$, is the task with the input/output pairs of $T$ inverted: $T^{\sim}\doteq\{ {\bf b} \rightarrow {\bf a} \}$. One requires that $(T^{\sim})^{\sim}=T$; and that $(T_1\otimes T_2)^{\sim}=T_1^{\sim}\otimes T_2^{\sim}$. 

A cardinal principle of constructor theory, called the {\sl composition law}, is that the composition of two possible tasks is a possible task. 

\subsubsection{A model of possible tasks and constructors in quantum theory}

I will now provide a quantum model for a constructor, following \cite{THE}. Consider the composite system of two quantum systems, $C$ and $S$, with total Hilbert space ${\cal H}={\cal H}_C\otimes {\cal H}_S$. Fix a unitary law of motion $U$ describing their interaction. I denote by $\Sigma(X)$ the +1-eigenspace of the projector $X$; also, for a general operator $B$, define $B^{(C)}=B\otimes \identity$ and $B^{(S)}= \identity\otimes B$. 
Fix two attributes of S, defined as ${\bf x}=\Sigma(X^{(S)})$ and ${\bf y}=\Sigma(Y^{(S)})$, where $X$ and $Y$ are two orthogonal projectors. Each of these attributes can be thought of as the set of states of $S$ in which the corresponding projector is sharp with value 1. Fix a task $T= \{{\bf x} \rightarrow {\bf y}\}$.

Consider now the set of states of $C$ defined as follows: $V_T=\{\ket{\psi}\in {\cal H}_C \;:\;\forall \ket{x}\in \Sigma(X^{(S)})\;,\;\;U(\ket{\psi}\ket{y})\in \Sigma({Y^{(S)}})\}.$ This is the set of states of $C$ with the property that, when $C$ is initialised in one of those states, and presented with the substrate $S$ in the state $\ket{x} \in \Sigma(G)$ with the attribute ${\bf g}$, it delivers the substrate in a state with attribute ${\bf y}$. Note that in the final state $C$ and $S$ can be entangled also, that $C$ may no longer be able to cause the transformation again once it has performed it once. 
\noindent It is straightforward to check that the $V_T$ is a vector space. A necessary set of conditions for $C$ to be a constructor for the task $A_t$ are: 
\begin{itemize}
\item{} $V_T$ is non empty;
\item{} There exists a subspace $W_T\subseteq V_T$ such that $$U\left (W_T\otimes \Sigma(X^{(S)})\right )\subseteq W_T\otimes \Sigma(Y^{(S)})\;.$$ These states of $C$ retain their property of being capable to cause the transformation $T$ over and over again. I shall denote by $\Pi_{W_T}$ the orthogonal projector onto the smallest subspace $W_T \subseteq {\cal H_C}$ with that property.
\end{itemize}
\noindent If the above two conditions are satisfied, we can define the (non-empty) set $V_{C_T}=\{\ket{\psi}\in {\cal H}_C \;:\;\forall \ket{x}\in \Sigma({X^{(S)}})\;,\;\;U(\ket{\psi}\ket{x})\in \Sigma({\Pi_{W_T}^{(C)}Y^{(S)}})\},$
which is easily proven to be a vector space. States in this subspace either belong to $W_{T}$ or they are brought into that space after one application of $U$. The projector $\Pi_{C_T}$ onto this subspace is the {projector for being a constructor} for the task $A_T$. 

That the task $T$ is possible implies in quantum theory that there exists a subspace $V_{C_T}$ with the above properties.

\subsubsection{Constructor Theory of Information}

Define the cloning task for the variable $X$ as:
\begin{equation}
\label{eq1}
C(X)\doteq\bigcup_{{x}\in X}\left\{({\bf x},{\bf x_0})\rightarrow ({\bf x},{\bf x}) \right\}\;
\end{equation}

\noindent {where $\bigcup$ is the set-union symbol and $x_0$ is a fixed attribute}. That a set $X$ is copiable means that the task $C(X)$ is possible, for some $x_0$. {In quantum theory, this task is possible whenever all elements in $X$ are orthogonal to one another; otherwise, if $X$ consists of non-orthogonal states, it is impossible. For example, when X is a boolean variable, $X=\{{\bf 0},{\bf1}\}$, and ${\bf x_0}={\bf 0}$, the task $C(X)$ can be implemented by a controlled-not gate.}  

An information medium is a substrate with the property that the cloning task $C(X)$ and the permutation task:

\begin{equation}
\label{eq2}
\Pi(X)\doteq \bigcup_{{x}\in X}\left\{{\bf x}\rightarrow \Pi({\bf x}) \right\}\;, 
\end{equation}

\noindent are all possible, for all permutations $\Pi$ on the set of labels of the attributes in $X$ and some attribute ${\bf x_0}\in X$. Once more, as an example, a set of orthogonal states in quantum theory, without additional symmetries or super-selection rules, qualifies as an information variable.

\noindent The task $C(X)$ corresponds to {\sl copying}, or cloning, the attributes of the first substrate onto the second, target, substrate; $\Pi(X)$, for a particular $\Pi$, corresponds to a logically reversible computation. For example, a qubit is an information medium with any set of two orthogonal quantum states, $X=\{{\bf 0}, {\bf 1}\}$, as defined above. 

As explained in the main text of this paper, a variable $Y$ is {\sl distinguishable} if the task

\begin{equation}
\label{eq3}
\bigcup_{{y}\in Y}\left\{{\bf y}\rightarrow {\bf q_y} \right\}
\end{equation}

\noindent is possible, where the variable $\{{\bf q_y}\}$, of the same cardinality as $Y$, is an information variable. 

Let me define $S(n)\doteq\underbrace{S\oplus S\oplus...S}_{n}$, a substrate consisting of n instances of the substrate $S$, and $x(n)\doteq\underbrace{(x,x,...,x)}_{n}$, attribute of  $S(n)$. Denote by $x(\infty)$ the attribute of $S(\infty)$, an unlimited supply of instances of $S$. Consider any two disjoint attributes $x$ and $x^{'}$. 

Asymptotic distinguishability requires the attributes $x(\infty)$ and $x^{'}(\infty)$ of $S(\infty)$, whenever they are defined, to be distinguishable {(hence, the variable $Y=\{x(\infty), x^{'}(\infty)\}$ is distinguishable as in \eqref{eq3})}.

\subsection{Appendix B: Conservation of energy} 

{I shall now express the requirement imposed by the law of conservation of energy in a scale- and dynamics-independent way, by formalising the observation that that a conservation law formulated by a given subsidiary theory induces a specific assignment of possible and impossible tasks on a physical system, \cite{DEU}. One can express this assignment with scale- and dynamics-independent statements, i.e. without appealing to any particular formalism such as Hamiltonian dynamics}. 

Consider the set $\Sigma$ of all tasks on a substrate $S$ consisting of only one input/output ordered pair. A conservation law for an additive quantity of the system $S$ (energy for instance) induces a partition of $\Sigma$ into equivalence classes, defined as follows. Each equivalence class $X_E$ has the property that for any two tasks $T_1$ and $T_2$ belonging to $X_E$:

\begin{itemize}
\item{} Either the tasks $\{T_1,T_2\}$ and their transposes $\{T_1^{\sim},T_2^{\sim}\}$ are all {\sl impossible}; or they are all possible. 
\item{} The task $T_1\otimes T_2^{\sim}$, and its transpose, are both possible tasks. 
\end{itemize}

{By using the properties of serial and parallel composition and the definition of transpose, \cite{MAT}, one can check that the two above conditions define an equivalence relation between tasks.} 

Using the properties of equivalence classes, one can introduce a real-valued function $F$, with the property that for any two pairwise tasks $T_1$, $T_2$, $F(T_1)=F(T_2)$ if and only if they belong to the same equivalence class.

{There are infinitely-many possible functions $F$ that could label the equivalence classes. How does one choose $F$ so that it expresses a given conservation law? By the properties of parallel and serial composition, one first notices that there is only one class where both $T_1$ and $T_2$ and their transposes are all possible. So, to express a conservation law with this approach, the key step is to select a function $F$ labelling the classes with the property that $F(T)=0$ for all tasks $T$ in that class. In all the other classes, any task $T$ and its transpose are both impossible: these classes can each be labelled by a non-zero value of $F(T)$. 
This is physically motivated as follows: upon this choice, the label $F(T)$ represents the amount by which the task $T$ violates the conservation law. In the class labelled by $F=0$, all tasks are possible as they do not require to modify the conserved quantity. In all the other classes, by our definition above, each task $T$ is impossible, but the task $T\otimes T^{\sim}$ is in turn possible.} So one can interpret the label $F(T)$ as the non-zero amount by which the task $T$ requires the conserved quantity to be changed: while $T$ is impossible, $T\otimes T^{\sim}$ is possible given that it requires to change $F$ by equal and opposite amounts on each substrate. 

Given that in this paper we are assuming for simplicity that the only conservation law is the conservation of energy, I shall call each equivalence class an {\sl energy-equivalence class}; if two tasks $T_1$ and $T_2$ belong to the same energy-equivalence class, I will write: $T_1\sim T_2$; which means that the two tasks $T_1$ and $T_2$ violate the conservation law, they do so by the same amount.  

Hence, the conservation of energy induces the scale-independent, dynamics-independent constraint that the possible and impossible tasks on substrates $S$ obey the two conditions listed above, with $F$ chosen as described. {The choice of the specific function $F$ and any further constraint on it are up to each particular dynamical theory to specify, and are not relevant for present purposes, because the theorems expressed in this paper hold at the meta-level of principles, which are intended to underlie all subsidiary theories that conform to them: from classical Hamiltonian mechanics and quantum theory, to other theories that we may yet have to discover.} 

By noticing that each task in $\Sigma$ is an ordered pair of attributes, the partition of tasks in $\Sigma$ into equivalence classes induces a partition into classes of the set of their input/output attributes. One can choose a function $E$ that labels each class by a real number, with the property that $E({\bf a})=E({\bf b})$ if and only if the two attributes belong to the same class, and if $T=\{{\bf a}\rightarrow{\bf b}\}$, then the function $F$ labelling the equivalence class of tasks is related to the function $E$ by the following relation: $F(T)=E({\bf b})-E({\bf a})$. The labelling of attributes defined by $E$ can be thought of, in this case, as an energy function (defined, as usual, up to a constant). Thus I shall say that an attribute has a particular value of energy if it belongs to one of these classes labelled by that particular value of energy, under a fixed labelling $E$ compatible with the partition into equivalence classes of the set of all pairwise tasks. It is also possible to show that the function $E$ has to be bounded both from above and from below, \cite{DEU, MAT}.  

\subsection{Appendix C: Theorem 1 in quantum theory}

\noindent In quantum theory, theorem 1 (see main section of this paper) can be proven by considering the general properties of programmable constructors - as outlined in \cite{THE}, which generalises a proof proposed in \cite{NIE}. I shall now summarise the key steps of the proof. 

Under laws of motion represented by a unitary $U$, a system $C$ may be a constructor for different tasks on the same substrate $S$ initialised to a generic, fixed attribute $G^{(S)}$ (that can be thought of as a blank state). For instance, let $\Pi_1$ be the projector for being a constructor for the task $A_{t_1}$ defined by the projector $T_1$, and $\Pi_2$ be the projector for being a constructor for the task $A_{t_2}$ associated with the projector $T_2$. In this case, $C$ can be considered as a programmable constructor with two kinds of programs in its repertoire, one to produce objects with the property $T_1$, the other to produce objects with the property $T_2$. (For example, $C$ could be the register of a quantum computer, $S$ its workspace.) Indeed, programs are (abstract) constructors.

\noindent Suppose that the two tasks are specified by unambiguous attributes, i.e, $\Sigma(T_1)\cap \Sigma(T_2)=\{0\}$. Then, one can prove that the projectors for the programs for each of those tasks must be orthogonal to each other: $\Pi_1\Pi_2=0$. 

\noindent By hypothesis, $U$ has the property that, for states $\ket{P_i}\in \Sigma(\Pi_i)$ 
\begin{eqnarray}
U(\ket{P_1}\ket{g})&\in& \Sigma({\Pi_1^{(C)}T_1^{(S)}})\\
U(\ket{P_2}\ket{g})&\in& \Sigma({\Pi_2^{(C)}T_2^{(S)}})\;.
\end{eqnarray}
No consider using the same program on several copies of the substrate $S^{(n)}=\underbrace{S\oplus S\oplus \dots \oplus S}_n$, each initialised in the legitimate input attribute: 
\begin{eqnarray}
\ket{P_1}\ket{g}^{\otimes n}&\rightarrow&\ket{\psi_1^{(n)}}\\ \nonumber
\ket{P_2}\ket{g}^{\otimes n}&\rightarrow&\ket{\psi_2^{(n)}}\nonumber
\end{eqnarray}
where $\ket{\psi_i^{(n)}}\in \Sigma \left(\Pi_i^{(C)}\hat T_i^{(n)}\right)$, where $\hat T_i^{(n)}= \identity \otimes \underbrace{T_i\otimes T_i \otimes \dots \otimes T_i}_n$. 
\noindent The above property must be true because, at the end of each transformation, the property of being a constructor for that specific task is preserved. 
\noindent Let us introduce the operator norm, $||A||=\displaystyle{{\rm Sup}\{|A\ket{v}|\;:|\ket{v}|=1\}}$. 
\noindent On the one hand this norm is a cross-norm: 
$||\hat T_1^{(n)}\hat T_2^{(n)}||= ||T_1T_2||^n$. 
\noindent On the other hand, $0\leq||T_1T_2||<1$ because the intersection 
between $\Sigma(T_1)$ and $\Sigma(T_2)$ is empty. This in 
turn follows from the theorem that the projector onto $\Sigma(T_1)\cup\Sigma(T_2)$ is 
$\lim_{n\rightarrow\infty}\left (T_1T_2\right)^n$. 

Which implies that, if the intersection is empty, 
there can be no non-zero states $\ket{v}$ with the property that 
$T_1T_2\ket{v}=\ket{v}$ (otherwise they would be in the 
intersection). This fact, together with the fact that $||T_1T_2||\leq||
T_1||\;||T_2||=1$ implies that $||T_1T_2||< 1$. 

\noindent Hence, in the limit 
$n\rightarrow \infty$ one has that $$|| \hat T_1^{(n)}\hat T_2^{(n)}||= || 
T_1T_2||^n \rightarrow 0$$ which implies that $$\lim_{n\rightarrow 
\infty}\hat T_1^{(n)}\hat T_2^{(n)}=0\;.$$

This means that the states $\lim_{n\rightarrow \infty}\ket{\psi_1^{(n)}}$, $\lim_{n\rightarrow \infty}\ket{\psi_2^{(n)}}$ are orthogonal, and so must be $\ket{P_1}$ and $\ket{P_2}$, because for arbitrary $n$ the transformation performed by that network is unitary. Picking the two pure states $\ket{P_1}$ in the $+1$-eigenspace of $\Pi_1$ and $\ket{P_2}$ in the $+1$-eigenspace of $\Pi_2$ with the property that $|\braket{P_1}{P_2}|^2$ is maximal, the above result shows that $|\braket{P_1}{P_2}|^2=0$, thus proving that $\Pi_1\Pi_2=0$. In other words, the network asymptotically works as a distinguisher between the two constructor subspaces.

\noindent Specialising this general result to the case analysed in the main section of this paper, one can simply take the attribute ${\bf w_0}$ appearing in the proof of theorem 1 to be $\Sigma (\Pi_1)$ and  ${\bf w_+}$ in that proof to be $\Sigma (\Pi_2)$, with $\Sigma (T_1)$ corresponding to $(\bf{w_+}, \bf{w_-})$, $\Sigma (T_2)$ corresponding to $(\bf{w_-}, \bf{w_+})$, and $\Sigma (G)$ corresponding to $(\bf{w_0}, \bf{w_0})$. Then the quantum-theory proof just outlined, showing that $\Pi_1\Pi_2=0$, implies that theorem 1 is true in quantum theory, as it implies that that the projector associated with the attribute ${\bf w_+}$ is orthogonal from the projector associated with the attribute ${\bf w_0}$, hence that the two attributes are distinguishable.

\end{document}